\documentstyle{article}
\newtheorem{theorem}{Theorem}[section]

\newenvironment{prooof}{\begin{description}
                   \item[{\small {\bf Proof:}}] \small}{\hfill {\bf Q.E.D.}
                                                          \medskip
                                                       \end{description}}

\newtheorem{defi}{Definition}[section]
\newtheorem{prop}{Proposition}[section]
\newtheorem{lemma}{Lemma}[section]
\newtheorem{rem}{Remark}[section]

\newcommand{\bdef}{\begin{defi}}
\newcommand{\ede}{\end{defi}}
\newcommand{\bsat}{\begin{theorem}}
\newcommand{\esat}{\end{theorem}}
\newcommand{\bprop}{\begin{prop}}
\newcommand{\eprop}{\end{prop}}
\newcommand{\blem}{\begin{lemma}}
\newcommand{\elem}{\end{lemma}}
\newcommand{\brem}{\begin{rem}}
\newcommand{\erem}{\end{rem}}
\newcommand{\bbew}{\begin{prooof}}
\newcommand{\ebew}{\end{prooof}}
\newcommand{\be}{\begin{equation}}
\newcommand{\ee}{\end{equation}}

\newcommand{\f}{\frac}
\newcommand{\p}{\partial}

\newcommand{\Real}{\mbox{I \hspace{-0.82em} R}}

\newcommand{\pbra}{\{ \hspace{-0.39em} \{ }
\newcommand{\pket}{\} \hspace{-0.39em} \} }

\begin{document}
\title{The Dynamics of Relativistic Membranes\\
         I: Reduction to 2-dimensional Fluid Dynamics \vspace{1cm}}
\author{{\bf Martin Bordemann}\\Fakult\"at f\"ur Physik\\Universit\"at
                             Freiburg\\
          Hermann-Herder-Str. 3\\79104 Freiburg i.~Br., F.~R.~G
        \and
        {\bf Jens Hoppe}\\Institut f\"ur Theoretische Physik\\Universit\"at
           Karlsruhe\\P.~O.~Box 6980\\76137 Karlsruhe, F.~R.~G.}
\date{FR-THEP-93-13 \\ KA-THEP-4-93 \\ June 1993}
\maketitle
\begin{abstract}
\noindent
We greatly simplify the light-cone gauge description of a relativistic
membrane moving in Minkowski space by performing a field-dependent change
of variables which allows the explicit solution of all constraints and a
Hamiltonian reduction to a $SO(1,3)$ invariant $2+1$-dimensional theory
of isentropic gas dynamics, where the pressure is inversely proportional
to (minus) the mass-density. Simple expressions for the generators of the
Poincar\'e group are given. We also find a generalized Lax pair which
involves as a novel feature complex conjugation. The extension to the
supersymmetric case, as well as to higher-dimensional minimal surfaces
of codimension one is briefly mentioned.
\end{abstract}
\vfill
\newpage

Who would believe that membranes are integrable?

\noindent On the other hand, realizing that geometry so often lies at the
heart of physics, it is also strange not to expect beautiful structures
and very special properties of relativistic surfaces, minimally embedded in
space-time.

The tribute for initiating into fruitful directions
concrete studies of relativistic membranes,
and higher-dimensional extended objects, should go to J.~Goldstone. This
`first phase' of membranes - not counting earlier ideas, such as
\cite{Dir62} - led to \cite{Hop82}. Five years later, starting with
\cite{KY86}, and the supersymmetrization of membranes (cf. \cite{HLP86}),
came a short blossom of activity (see for instance \cite{proc}), bringing
also into fashion $sdiff$ algebras, their star
product deformations, and their approximations via nontrivial large $N$
limits of matrix algebras (cf. e.~g. \cite{BHSS91}). The interest in
membranes more or less stopped, around 1990, for presumably two reasons:
The fact that `nothing could be solved' (not even on the classical level),
together with apparent proofs of nonintegrability (\cite{NSZ};
note \cite{DR90}), and
arguments (based on the large $N$ matrix model regularisation of
the membrane, cf. \cite{Hop82}, \cite{dHN88}) that the supersymmetric
membrane Hamiltonian
should have a continuous spectrum starting at zero (cf. \cite{dLNS}).

In contrast to all these negative results, and speculations, we would like
to put forward, in a series of papers devoted to different
aspects of relativistic surfaces (starting with the classical
dynamics of a bosonic membrane moving in four-dimensional flat Minkowski
space)
a more optimistic point of view, which is mainly based on two observations:

Firstly, one may explicitly solve the $sdiff$-constraint which
so far seemed to be untractable.

Secondly, the theory is greatly simplified by performing a time-dependent
change of variables, that interchanges some of the dependent and
independent variables.

Both these observations use in an essential way the infinite dimensionality
of the phase space, putting further doubt on conclusions drawn from the
finite-dimensional regularisation of the membrane.

That hodograph-type transformations play an important r\^ole in the
investigation
and solution of nonlinear dynamics has a long history while having become
particularly clear (\cite{TsBWTFH}) in the context of
{\em higher dimensional} integrability.

The paper is organised as follows:
After a brief review of light-cone gauge membrane dynamics
(see \cite{Hop82} or \cite{Hop87} for details) we show that
a change from the independent to the dependent variables allows us to
solve (actually integrate) the constraint and to reduce the original
system of field equations, which involves four functions, to a system
involving
only two functions, $q$ and $p$. The reduction is Hamiltonian, and $q$ and
$p$
can be interpreted as a gas-dynamical mass density and a velocity
potential,
respectively. We then write down a (complex conjugated) Lax equation for
the system, and finally give explicit formulae for the infinitesimal
generators of the Poincar\'e group in this model.
\vspace{0.5cm}
\newline

{\bf 1.} Let $\Sigma$ be a two-dimensional surface
with co-ordinates $(\varphi_1,\varphi_2)=\vec{\varphi}$ with respect to
which there is a Poisson bracket for any two smooth complex-valued
functions $f$ and $g$ on $\Sigma$:
\be \{f,g\} \stackrel{\rm def}{=}
   \f{\p f}{\p \varphi_1} \f{\p g}{\p \varphi_2}
       -\f{\p f}{\p \varphi_2} \f{\p g}{\p \varphi_1} \hspace{2cm}.
\ee
In the orthonormal light-cone gauge, the dynamics of a bosonic membrane
$x^{\mu}: \Real \times \Sigma \rightarrow \Real^{(1,3)}$ in
four-dimensional
Minkowski space is described by the two transverse co-ordinate functions
$(x_1,x_2)=\vec{x}$ and their conjugate momenta $(p_1,p_2)=\vec{p}$,
whereas one of the light-cone co-ordinates, $\zeta$, becomes
cyclic (such that its conjugate momentum $-\eta$ is conserved).
$\eta$ times the Hamiltonian governing the dynamics of the membrane
is then given by
\be \label{ham}
H~=~\f{1}{2} \int_{\Sigma} d\varphi_1d\varphi_2(p_1^2+p_2^2+\{x_1,x_2\}^2)
                                    \hspace{2cm}, \ee
\noindent provided the constraint
\be \label{con} K(\vec{\varphi})~\stackrel{\rm def}{=}~
            \{p_1,x_1\}(\vec{\varphi})+\{p_2,x_2\}(\vec{\varphi})
                ~\stackrel{!}{=}~0 \ee
\noindent holds which reflects the remainder of the
diffeomorphism invariance of the original relativistic action (that was
taken to be proportional to the three-dimensional volume swept out in
Minkowski space). Using the usual dynamical Poisson bracket
$\pbra~~,~~\pket $ on the classical fields,
\be \pbra x_i(\vec{\varphi}),p_j(\vec{\varphi'})\pket ~=~\delta_{ij}
                         \delta^{(2)}(\vec{\varphi}-\vec{\varphi'})
                                       \hspace{1cm}, \ee
\noindent one gets the following equations of motion from the Hamiltonian
(\ref{ham}):
\begin{eqnarray}
  \dot{x_1} & = & p_1 \label{x1} \\
  \dot{x_2} & = & p_2 \label{x2} \\
  \dot{p_1} & = & \{ \{x_1,x_2\},x_2\} \label{p1} \\
  \dot{p_2} & = & - \{ \{x_1,x_2\},x_1\} \label{p2}
\end{eqnarray}
\noindent which imply
\be \label{x12} \dot{\{x_1,x_2\} }~=~\{p_1,x_2\}+\{x_1,p_2\} \hspace{2cm}.
\ee
\vspace{0.5cm}
\newline

{\bf 2.} In order to obtain an unconstrained description of the membrane
we make the following change of independent variables:
\be \label{tr} (\varphi_0 \stackrel{\rm def}{=} t, \varphi_1, \varphi_2)
       \mapsto (x_0=t, x_1(t,\vec{\varphi}), x_2(t,\vec{\varphi}))
                            \hspace{2cm}, \ee
\noindent with the Jacobian
\be \label{jac} (\f{\p \vec{x}}{\p \vec{\varphi}})
         ~=~\left( \begin{array}{ccc}
            1      &         0             &          0            \\
         \dot{x_1} & \f{\p x_1}{\p \varphi_1} & \f{\p x_1}{\p \varphi_2} \\
         \dot{x_2} & \f{\p x_2}{\p \varphi_1} & \f{\p x_2}{\p \varphi_2}
              \end{array} \right)  \ee
\noindent and its Jacobi determinant
\be J(\vec{\varphi}) \stackrel{\rm def}{=}
         \det(\f{\p\vec{x}}{\p\vec{\varphi}})(\vec{\varphi})
               ~=~\{x_1,x_2\}(\vec{\varphi}) \hspace{2cm}. \ee
Along with this transformation (assumed to be nonsingular)
we use the notation $f(\vec{\varphi})=\hat{f}(\vec{x})$ for any function
$f$ on $\Sigma$ to indicate the transformation.

This transformation is tailored to simplify Poisson brackets with
the fields $x_i$ ($i=1,2$; $\epsilon_{12} = -\epsilon_{21} = 1$):
\be \{f,x_i\}(\vec{\varphi})~=~-\epsilon_{ij} \f{\p \hat{f}}{\p x_j}
                              (\vec{x})
                                 \hat{J}(\vec{x}) \hspace{2cm}.\ee

An immediate consequence is a simplification of the
constraint equation (\ref{con}):
\be \label{connew} K(\vec{\varphi})
        ~=~(\vec{\nabla} \times \hat{\vec{p}})(\vec{x})\hat{J}(\vec{x})~=~0
\ee
where $\vec{\nabla} \times \hat{\vec{p}}\stackrel{\rm def}{=}
\p_1\hat{p_2}-\p_2\hat{p_1}$.
The solution of eqn (\ref{connew}) is simply
a gradient of some function $p$ depending on the $\vec{x}$-variables:
\be \hat{\vec{p}}=\vec{\nabla}p \ee
Actually, we had first solved the constraint in the original co-ordinates:
\begin{eqnarray} p_1 & = & \f{\{P,x_2\}}{\{x_1,x_2\}} \\
                 p_2 & = & -\f{\{P,x_1\}}{\{x_1,x_2\}} \hspace{2cm},
\end{eqnarray}
the function $p$ being equal to $\hat{P}$.
The verification that this indeed satisfies the original constraint
(\ref{con}) is simple if one uses the fact that in two dimensions every
Poisson bracket remains a Poisson bracket when multiplied by an arbitrary
smooth function.
Note again that this uses in an essential way the infinite
dimensionality of the phase space and does not seem to have any immediate
analogue in the large $N$ matrix model regularisation of the membrane.

Now the equations of motion can be reduced:
Using (\ref{jac}) one observes that the `old' time derivative
becomes a substantial (or material) derivative known from fluid dynamics:
\be \f{\p}{\p \varphi_0}=(\vec{\nabla}\cdot \vec{p}) \cdot \vec{\nabla}
                        + \f{\p}{\p x_0}  \hspace{2cm}.  \ee
The r.~h.~s. of eqs (\ref{p1}, \ref{p2}) becomes an
ordinary {\em first} derivative (of $\hat{J}^2$) in the
$\vec{x}$-co-ordinates.
Now denoting by $(\dot{~~})$ the differentiation with respect to $t=x_0$
and defining
\be q(\vec{x}) \stackrel{\rm def}{=} \f{1}{\hat{J}(\vec{x})}
                  = \det(\f{\p \vec{\varphi}}{\p \vec{x}})(\vec{x}) \ee
we get
\begin{eqnarray}
 \dot{q} & = & -q(\vec{\nabla})^2p-\vec{\nabla}p
                          \cdot \vec{\nabla}q  \label{Jeqn} \\
 \dot{p} & = & \f{1}{2}(\f{1}{q^2}-(\vec{\nabla}p)^2)+c(t)  \label{peqn}
\end{eqnarray}
where $c(t)$ arises
as an integration `constant' (for the moment, we will take it to be zero).
What about eqs (\ref{x1}, \ref{x2}) which were used to obtain eqn
(\ref{x12})? They are equivalent to the equations
\be (\f{\p}{\p x_0}+(\vec{\nabla}p)\cdot \vec{\nabla})\varphi_i~=~0
                              \hspace{2cm}.\ee

Taking the gradient w.~r.~t. $\vec{x}$ of
eqn (\ref{peqn}) we get a pair of eqs that are well-known in fluid
dynamics:
\begin{eqnarray}
    \dot{q}+\vec{\nabla}\cdot(q\vec{\nabla}p) & = & 0 \\
   q\dot{\vec{\nabla}p}+q(\vec{\nabla}p\cdot\vec{\nabla})\vec{\nabla}p
               +\vec{\nabla}(-\f{1}{q}) & = & 0 \hspace{2cm}.
\end{eqnarray}
Viewing $q$ as a mass density and $\vec{v}=\vec{\nabla}p$ as a velocity
field
the first equation is nothing but the continuity equation whereas
the second one is the Euler equation of irrotational
($\vec{\nabla}\times \vec{v}=0$) isentropic 2-dimensional gas dynamics with
a pressure $-\f{1}{q}$. The function $p$ plays the r\^ole of a
velocity potential.

Another feature of the reduced eqs (\ref{Jeqn}, \ref{peqn}) is that they
are again {\em Hamiltonian}: treating $q$
and $p$ as canonical variables, i.~e.
\be  \pbra q(\vec{x}),p(\vec{y})\pket =\delta^{(2)}(\vec{x}-\vec{y}) \ee
and expressing the old Hamiltonian (\ref{ham}) in the reduced variables
\be \label{newh}
 \tilde{H}~=~\f{1}{2}\int \! d^2x~q(\vec{x})((\vec{\nabla}p(\vec{x}))^2+
                                             \f{1}{q(\vec{x})^2})  \ee
the above equations of motion can be deduced from $\tilde{H}$ in the
canonical way, a fact which does not seem to be mentioned in textbooks
on gas dynamics.

{}From the point of view of phase space reduction these features are quite
natural:
the left hand side of the constraint equation (\ref{con})
can be viewed as a momentum map for the $SDiff(\Sigma)$ (i.~e. group of
area
preserving diffeomorphisms on $\Sigma$) action on the original fields
$\vec{x}$.
Therefore we did a phase space reduction at momentum level zero. One
could also consider a reduction at a nonzero momentum level thus
introducing
a fixed degree of vorticity in the reduced phase space living on additional
co-adjoint orbits of $SDiff(\Sigma)$.
It is interesting that this analysis
does not depend on the fact that $\Sigma$ is two-dimensional (we shall give
a more detailed differential geometric account thereof elsewhere).

When extending the analysis to the supersymmetric case, one finds (with
$\psi(\vec{x})$ a complex anti-commuting field)
\begin{eqnarray}
   \tilde{H}_s & = & \f{1}{2}\int \! d^2x~\big[ q(\vec{\nabla}p)^2
                  +{\bf i}p(\psi \vec{\nabla}^2\bar{\psi}
                            +\bar{\psi}\vec{\nabla}^2\psi) \nonumber \\
               &   & \hspace{0.4cm}+\f{1}{q}(1
                        +\f{\psi\bar{\psi}}{4}
                        (\p\bar{\psi}\bar{\p}\psi-\p\psi\bar{\p}\bar{\psi})
                        +{\bf i}(\psi\bar{\p}\psi+\bar{\psi}\p\bar{\psi}))
                      \big]  \hspace{0.6cm}.
\end{eqnarray}
Its derivation and canonical structure (which is more
complicated than in the purely bosonic case)
will be given in a separate paper.

Note that the connection between eqn (\ref{ham}) and eqn (\ref{newh})
has, as a simple consequence, the solution of constant pressure
irrotational gas dynamics
\be \label{newh0}
  \tilde{H_0}~=~\f{1}{2}\int \! d^2x~q(\vec{x})(\vec{\nabla}p(\vec{x}))^2
\ee
in terms of `free' fields $(x_1(\vec{\varphi}),x_2(\vec{\varphi}))$
only subjected to the constraint (\ref{con}):
Just solve the equations in the old variables, i.~e.
$\ddot{x_i}=0$ for $i=1,2$ and $\{x_1,\dot{x_1}\}+\{x_2,\dot{x_2}\}=0$,
trivially giving
\be
x_i(t,\vec{\varphi})~=~x_i(0,\vec{\varphi})+t\dot{x_i}(0,\vec{\varphi})
                                     \label{dyn0} \ee
\be \{x_1(0,\cdot),\dot{x_1}(0,\cdot)\}+\{x_2(0,\cdot),\dot{x_2}(0,\cdot)\}
                   ~=~0  \hspace{1cm}.\ee
For any choice of the initial conditions eqn (\ref{dyn0}) when
inverted to functions $\varphi_i(t,\vec{x})$ will give $q(t,\vec{x})$ as
the
Jacobi determinant $\det(\p\varphi_i/\p x_j)$. This is a nice (infinite
dimensional!) example of the projection method of Olshanetsky and Perelomov
(cf. \cite{OP81})

Finally note that the one-dimensional analogue of the Hamiltonian
(\ref{newh})  can presumably be solved by quite a variety of different
methods such as direct linearization, infinite charge algebra, collective
field theory or matrix model limit.
\vspace{0.5cm}
\newline

{\bf 3.} Going back to the membrane Hamiltonian (\ref{newh}) observe that
the equations of motion take a remarkably simple form - exhibiting the
special nature of the $\f{1}{q^2}$-potential - when using the linear
differential operator
\be D \stackrel{\rm def}{=} q(\f{\p}{\p t}
                             +(\vec{\nabla}p)\cdot\vec{\nabla})   \ee
\noindent and complex co-ordinates $z=\f{1}{2}(x_1+{\bf i}x_2)$,
$\p=\p_1-{\bf i}\p_2$, $\bar{\p}=\p_1+{\bf i}\p_2$, namely:
\begin{eqnarray}
               D(\f{1}{q}) & = & \p \bar{\p}p \\
                   D(\p p) & = & \p (\f{1}{q})  \\
              D(\bar{\p}p) & = & \bar{\p}(\f{1}{q}) \hspace{1cm}.
\end{eqnarray}
Defining the following functional $l$ of the fields
\be \label{lax} l~=~\f{e^{-{\bf i}\theta}}{q}+e^{{\bf i}\theta}\p p  \ee
with $e^{{\bf i}\theta}$ a spectral parameter (reflecting the explicit
$U(1)$-invariance of the model) the equation
\be \label{laxen} Dl=\p \bar{l}  \ee
encodes the membrane-dynamics. Since the integral
$\int \! d^2\varphi$ over the $\varphi$-variables becomes an integral
$\int \! d^2x~q(\vec{x})$ (with density $q$) under the
transformation (\ref{tr}) it is clear that if
\be \int \! d^2x~ D(C(q,p,\p p,\bar{\p}p,...)~=~0  \ee
for some functional $C$ of the fields $q$ and $p$, the
integral over the $\varphi$-variables of the corresponding functional in
the original variables will be a constant of motion. Moreover, it is
easy to check that eqns (\ref{lax}) and (\ref{laxen}) can be retransformed
as follows:
\be L \stackrel{\rm def}{=}
    e^{{\bf i}\theta}\{x_1,x_2\}+e^{{\bf i}\theta}(p_1-{\bf i}p_2) \ee
satisfies the `Lax like' equation
\be \label{laxeq} \dot{L} = \{\bar{L},M\}  \ee
(where the function $M$ is given by $M={\bf i}e^{-{\bf i}\theta}
(x_1-{\bf i}x_2$)) {\em provided the constraint holds}. The original
Hamiltonian (\ref{ham}) is simply given by
\be H~=~\f{1}{2}\int \! d^2\varphi~ \bar{L}L
       ~=~\f{1}{2}\int \! d^2x~ \bar{l}l
             \hspace{1cm}.\ee
Note that the $\theta$-dependent cross terms automatically
vanish upon integration due to the constraint eqn (\ref{con}).
\vspace{0.5cm}
\newline

{\bf 4.} Let us now comment on the 4-dimensional Poincar\'e invariance of
the membrane : Consider the following functionals of the fields $q$ and $p$
(where $i=1,2$):
\begin{eqnarray}
P_+ & \stackrel{\rm def}{=} & \eta \label{gen1} \\
P_- & \stackrel{\rm def}{=} & \f{\tilde{H}}{\eta} ~=~
 \f{1}{2\eta}\int \! d^2x~q((\vec{\nabla}p)^2+\f{1}{q^2}) \label{gen2} \\
P_i & \stackrel{\rm def}{=} & \int \! d^2x~ q\p_i p    \label{gen3} \\
J_{12} & \stackrel{\rm def}{=} & \int \! d^2x~ q(x_1\p_2 p-x_2\p_1 p)
                                                         \label{gen4} \\
J_{i+} & \stackrel{\rm def}{=} & \eta \int \! d^2x~ qx_i  \label{gen5} \\
J_{+-} & \stackrel{\rm def}{=} & -\eta \zeta_0  \label{gen6}  \\
J_{i-}~& \stackrel{\rm def}{=} & \f{1}{2\eta}\int \! d^2x~qx_i
           ((\vec{\nabla}p)^2+\f{1}{q^2})
           ~-~\f{1}{2\eta}\int \! d^2x~q\p_i(p^2) \nonumber \\
          & &  ~-~\f{1}{\int \! d^2x~ q}
               (\zeta_0-\f{\int \! d^2x~qp)}{\eta})\int \! d^2x~q \p_ip
                \label{gen7}     \hspace{1cm}.
\end{eqnarray}
Here, $\zeta_0$ is the integral over the light-cone co-ordinate $\zeta$ of
the membrane. Simply using the canonical Poisson brackets
\begin{eqnarray}
{}~ \pbra q(\vec{x}),p(\vec{x'})\pket  & = & \delta^{(2)}(\vec{x}-\vec{x'})
                                                                  \\
{}~ \pbra\zeta_0,\eta\pket  & = & -1
\end{eqnarray}
one may convince oneself that the $J$'s above constitute a basis of the
Lie algebra $so(1,3)$ of the Lorentz group, and that the $P$'s commute
among themselves and with the $J$'s in the correct way. In particular
($i=1,2$):
\begin{eqnarray}
{}~ \pbra J_{i-},P_j\pket  & = & \delta_{ij}P_-  \\
{}~ \pbra J_{i-},P_+\pket  & = & P_i             \\
{}~ \pbra J_{i-},P_-\pket  & = & 0               \\
{}~ \pbra J_{1-},J_{2-}\pket  & = & 0      \\
{}~ \pbra J_{i+},J_{k-}\pket  & = & \delta_{ik}J_{+-}-J_{ik} \hspace{2cm}.
\end{eqnarray}
The relativistic mass of the membrane
\be {\rm M}^2~~=~~2P_+P_--\vec{P}^2~~=~~
       \int \! d^2x~q((\vec{\nabla}p)^2+\f{1}{q^2})~-~P_1^2~-~P_2^2 \ee
Poisson-commutes with all the ten generators given in eqs. (\ref{gen1}) -
(\ref{gen7}).

\end{document}